\documentclass{ws-procs9x6}            
\usepackage{graphicx}
\begin{document}
\title{Determination of measurement uncertainty by Monte Carlo simulation}

\author{Daniel Hei{\ss}elmann$^*$, Matthias Franke, Kerstin Rost, and Klaus Wendt}

\address{Coordinate Metrology Department, Physikalisch-Technische Bundesanstalt (PTB), Bundesallee 100,
Braunschweig, 38116, Germany\\
$^*$E-mail: daniel.heisselmann@ptb.de}

\author{Thomas Kistner}

\address{Carl Zeiss Industrielle Messtechnik GmbH, Carl-Zeiss-Stra{\ss}e 22,\\
73446 Oberkochen, Germany}

\author{Carsten Schwehn}

\address{Hexagon Metrology GmbH, Siegmund-Hiepe-Stra{\ss}e 2-12,\\
35578 Wetzlar, Germany}

\begin{abstract}
Modern coordinate measurement machines (CMM) are universal tools to measure geometric features of complex three-dimensional workpieces. To use them as reliable means of quality control, the suitability of the device for the specific measurement task has to be proven. Therefore, the ISO 14253 standard requires, knowledge of the measurement uncertainty and, that it is in reasonable relation with the specified tolerances. Hence, the determination of the measurement uncertainty, which is a complex and also costly task, is of utmost importance.\par
The measurement uncertainty is usually influenced by several contributions of various sources. Among those of the machine itself, e.g., guideway errors and the influence of the probe and styli play an important role. Furthermore, several properties of the workpiece, such as its form deviations and surface roughness, have to be considered. Also the environmental conditions, i.e., temperature and its gradients, pressure, relative humidity and others contribute to the overall measurement uncertainty.\par
Currently, there are different approaches to determine task-specific measurement uncertainties. This work reports on recent advancements extending the well-established method of PTB's Virtual Coordinate Measuring Machine (VCMM) to suit present-day needs in industrial applications. The VCMM utilizes numerical simulations to determine the task-specific measurement uncertainty incorporating broad knowledge about the contributions of, e.g., the used CMM, the environment and the workpiece. 
\end{abstract}


\bodymatter

\section{Introduction}
Knowledge about the measurement uncertainty is a basic requirement for quality-driven and economic production processes and involves the entire production technique. 
However, in coordinate metrology the determination of measurement uncertainties is a complex process that has to be performed task-specific. Particularly, size, shape, form deviations, and accessibility of the feature significantly influence the achievable measurement uncertainty. 
Thus, the large spectrum of parts with several variations and narrow tolerances at the same time require methods to determine the measurement uncertainty in a simple and efficient, yet universal manner. 
Generally, there are three different approaches for the calculation of the measurement uncertainty: analytical budgets, experimental determination, and numerical simulations.\par
All three require suitable mathematical models as well as the description and quantification of uncertainty contributions. The uncertainty determination by an analytical uncertainty budget based on detailed knowledge of all individual contributions is described by the GUM \citep{GUM_EN,sommer_siebert2004tm}. 
 The experimental approach according to ISO~15530-3\cite{ISO_15530-3:2011} demands a calibrated workpiece and multiple measurements. ISO/TS~15530-4\cite{ISO_TS_15530-4:2008-06} describes general requirements for the application of simulation methods, like the one presented in this work.


\section{Measurement uncertainty determination by numerical simulation}
\begin{figure}[tbp]
	\center
	\includegraphics[trim = 0mm 0mm 0mm 0mm, clip, width=0.50\columnwidth]{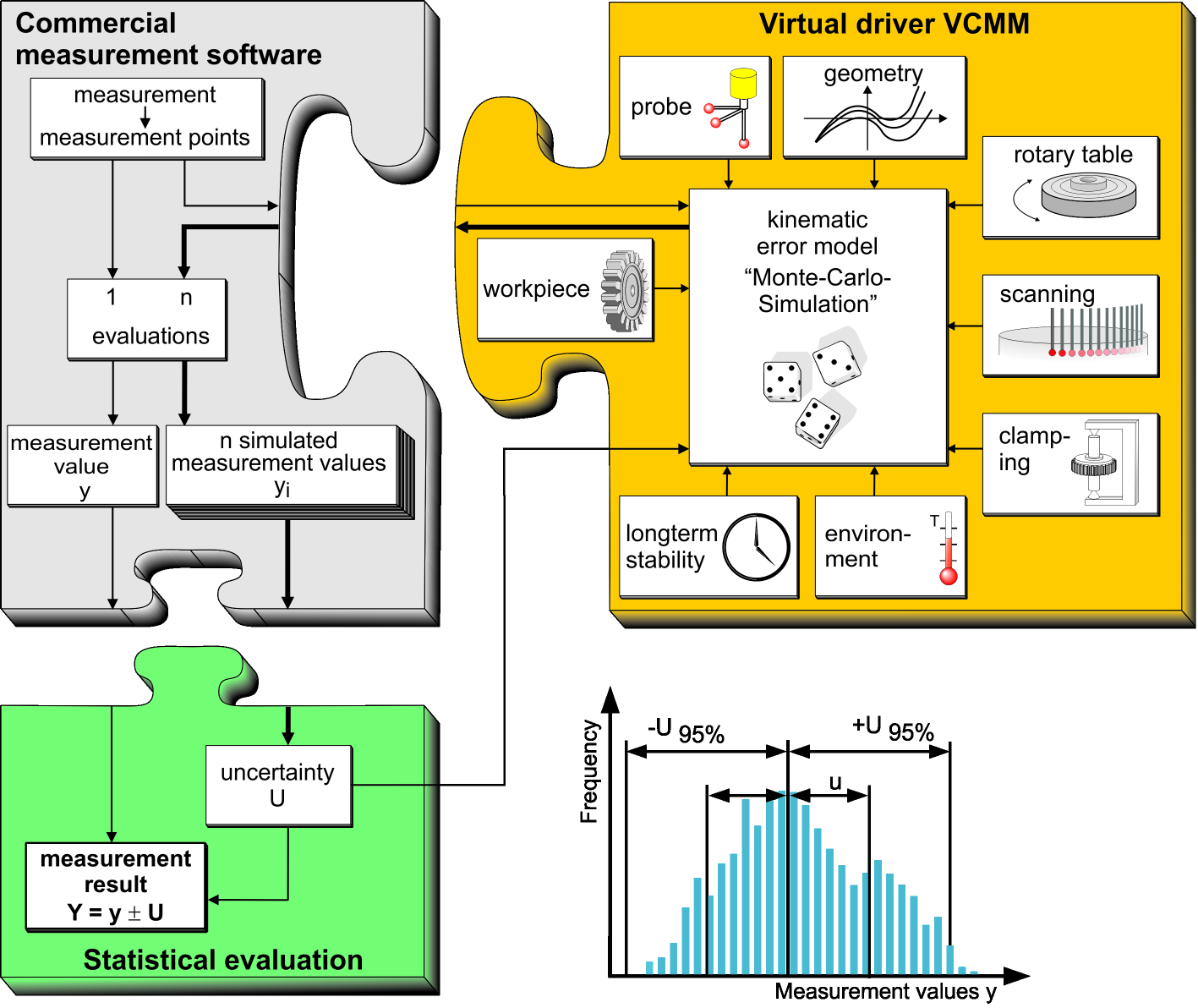}
	\caption{Principle sketch of the virtual coordinate measuring machine VCMM: The CMM software passes the measured values to the VCMM.  The statistical analysis is performed by a separate module.}
	\label{f_puzzle}
\end{figure}

Numerical simulations 
are an efficient and versatile technique to universally determine the measurement uncertainty in coordinate measurements. The basic elements of the realization of this method are described by GUM Supplement 1 \cite{GUM_Supp1_EN}. The ``Virtual Coordinate Measuring Machine'' VCMM 
is a software tool that implements the simulation method. It is based on an approach to consider input parameters for all influences occurring during the measurement procedure, including those that are not machine specific (Fig. \ref{f_puzzle}). All parameters are described by suitable probability distribution functions and are linked to mathematical models evaluating their effect on the measurement of an individual position. 
The evaluation of the measurement uncertainty is performed in a six-step process, where (1) and (2) 
are the same as in any measurement:
\newcounter{VCMM}
\begin{enumerate}
	\item The CMM records the measurement points of all probed features of the workpiece.
	\item The CMM's analysis software calculates the features' characteristics (i.e., measurement values) from the recorded data points.
	\setcounter{VCMM}{\value{enumi}}
\end{enumerate}
After these tasks are accomplished, four additional steps are required:
\begin{enumerate}
	\setcounter{enumi}{\value{VCMM}}
	\item Based on the measured coordinates, the  VCMM software calculates a set of new coordinates that have been modified considering systematical and random measurement deviations.
	\item These coordinates are  analyzed in the same way as the original coordinates.
	\item Steps (3) and (4) are repeated $n$ times until a stable calculation of the measurement uncertainty has been achieved. 
	\item The measurement results and their uncertainties are calculated from the distributions and the mean values of all $n$ repeated simulations.
\end{enumerate}
The first version of PTB's VCMM was published 15 years ago \citep{schwenke_waeldele2002tm}. Since then, the software has been thoroughly revised, modularized and implemented into object-oriented code. The mathematical models have been adapted to the technical advancement of CMM, wherever necessary. Furthermore, advanced algorithms for the creation of random numbers \citep{rost2016PhD_EN} and 
the analysis of scanning data and consideration of undersampling of features have been implemented. Additionally, the VCMM was augmented by a statistics module, checking the stability of the computation of the measurement uncertainty according to predefined criteria. Depending on the specific application 
the VCMM offers different opportunities to choose the input parameters. In order to achieve the highest precision it is recommended to precisely determine the machine-specific measurement deviations after commissioning the CMM. 
Alternatively, the machine-specific input parameters may be determined using manufacturer's specifications, such as maximum permissible errors (MPE).
\par
\begin{figure}[tbp]
	\center
	\includegraphics[trim = 0mm 0mm 0mm 0mm, clip, width=0.5\columnwidth]{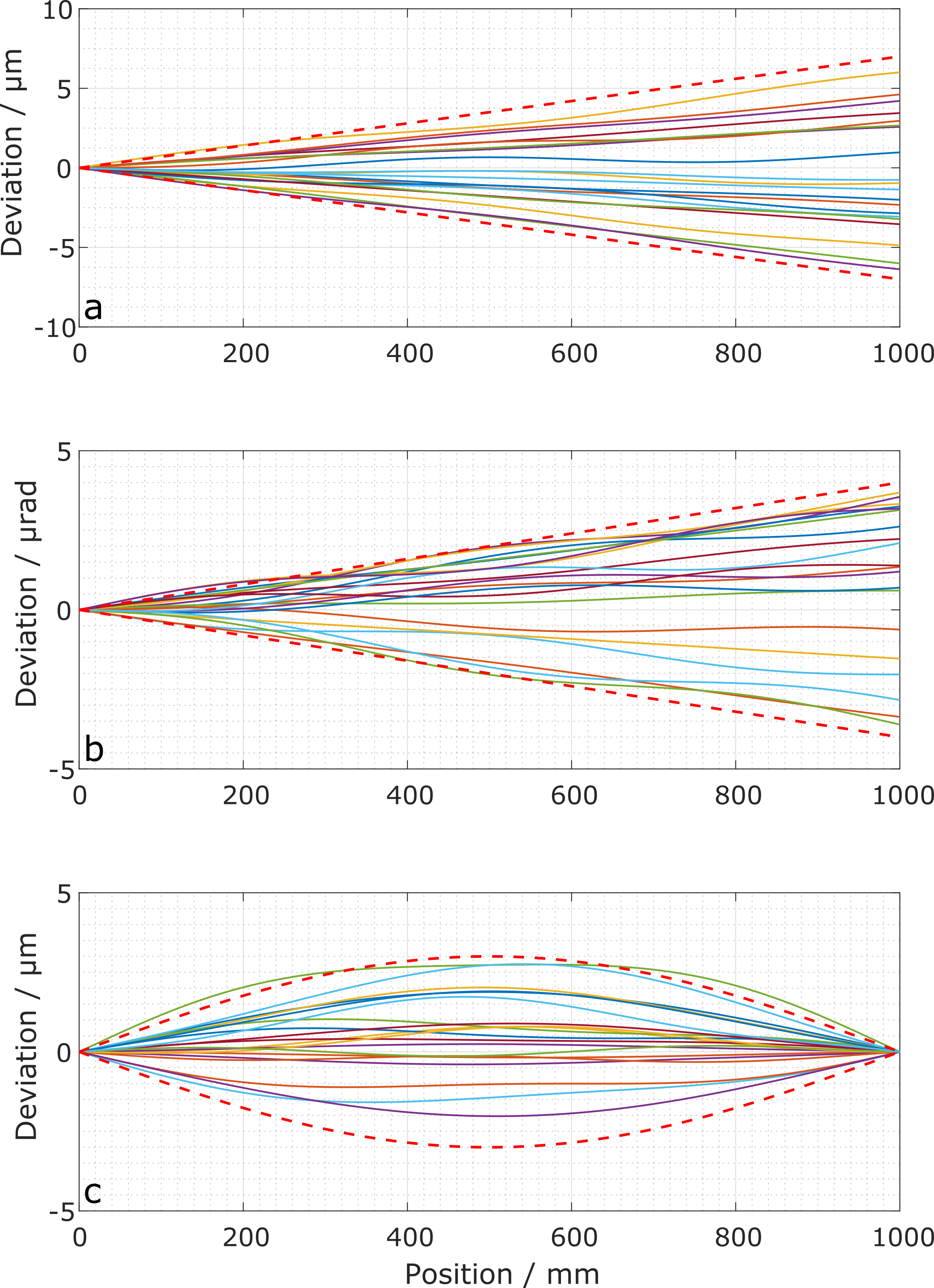}
	\caption{Example of machine-specific input parameters from manufacturer's specifications. The solid curves represent a set of simulated deviations in position (a), rotation (b), and straightness (c) within the maximum permissible errors (MPE) specified by the CMM manufacturer (dashed curves).}
	\label{f_ErrorFunctions}
\end{figure}

 
The simple parametrization of the VCMM does not consider detailed characteristics of the individual parameters in the same way as a complete determination of residual errors of the individual CMM (Fig. \ref{f_ErrorFunctions}). Therefore, the measurement uncertainties derived for a machine characterized by threshold values 
will in general be larger than those of a CMM that was individually parameterized using calibrated standards.

\section{Input parameter models}
The VCMM software is capable of calculating the measurement uncertainty for measurements of individual characteristics on CMMs. Therefore, the it takes into account \emph{(i)} geometry errors of the guideways, \emph{(ii)} measurement errors of the probing system for recording individual data points or scanning measurements, \emph{(iii)} thermally-induced deformations of the CMM structure, \emph{(iv)} influences of the workpiece
, and \emph{(v)} error contributions of the measurement strategy.
The individual error components treated in the VCMM are comprised of up to three different types of  contributions:
\begin{itemize}
	\item A known and systematic contribution that is not varied during the runtime of the VCMM software 
	imposes a systematic bias on the measurement results.
	\item A stochastic contribution, which is altered before each VCMM run or on the occasion of certain events. 
	 It is varied depending on the specified uncertainties and reflects residual systematic errors.
	\item Entirely random measurement uncertainty contributions are independently varied for each measurement point.
\end{itemize}
The aforementioned contributions depend on the positions $\vec{p}$ of the CMM guideways, the stylus tip offset $\vec{l}$ and also the probing direction $\vec{n}$ during recording of a measurement point. They can either be described as a constant value that is considered independent of the position, a length-dependent contribution, linearly increasing with increasing length of travel, a linear combination of several harmonic functions, or look-up table. In general, these options can also be combined. \par

\subsection{Guideway geometry}
The static guideway geometry errors are treated by the VCMM based on a rigid body model. The errors are determined from the translational and rotational deviations of the guideways, their respective squareness deviation, the uncertainty of the length determination as well as short-periodic deviations of guideways and scales. The three axes of the CMM are considered as independent and the results are obtained by superposition of the various contributions.
The position error $\vec{e}_{G_i}$ resulting from static guideway errors can thus be described by 
\begin{eqnarray}
\vec{e}_{G_i} = & \vec{e}_{GT_i} + \mathbf{E_{GR_i}}\cdot \vec{p}_i + \mathbf{E_{GP_i}}\cdot \vec{l} +(1+s_{GM})\cdot \vec{p}_i \;,\label{eq_statgeom}
\end{eqnarray} 
where $\vec{e}_{GT_i}$ is the translational component, $\mathbf{E_{GR_i}}\cdot \vec{p}_i $ resembles the rotational contributions, $\mathbf{E_{GP_i}}\cdot \vec{l}$ comprises the contributions of the probing element, and the term $(1+s_{GM})\cdot \vec{p}_i$ adds the uncertainty of the length measurement.
Within the VCMM software the error components are concatenated using a systematic and a stochastically changing contribution, where the latter is evaluated during the VCMM's runtime according to the specified uncertainties.\par

\subsection{Thermal deformation of the machine}
Besides the CMM's static geometry errors, additional contributions result from thermal effects, such as spatial and temporal temperature gradients or variations of the mean temperature. The 
 information about the strength of these effects is generally obtained from long-term observations of the CMM and its environment. The implementation of the VCMM considers the thermal expansion of the scales and the guideways as well as the thermally-induced straightness errors, rotational errors, and squareness errors. 
 Besides changes in the mean temperature of the measurement environment, the variation of thermal gradients within the CMM and the measurement room and the uncertainty of the temperature measurement are considered.
With respect to the thermal expansion of the scales, the VCMM distinguishes between CMM with and without temperature compensation. For the latter, the deviation from the mean temperature and temperature of the scale are considered to contribute proportional to both the length of travel and the coefficient of thermal expansion. The uncertainty of the determination of the thermal expansion coefficient requires the addition of another term. Following (\ref{eq_statgeom}), the uncertainty arising from thermally-induced deformation can be described by the model equation
\begin{eqnarray}
\vec{e}_{T_i} = & \vec{e}_{TT_i} + \mathbf{E_{TR_i}}\cdot \vec{p}_i + \mathbf{E_{TP_i}}\cdot \vec{l}\;.\nonumber \label{eq_thermdeform}
\end{eqnarray} 
The vector $\vec{e}_{TT_i}$ combines the thermally-induced straightness deviations, while the terms $\mathbf{E_{TR_i}}\cdot \vec{p}_i $ and $\mathbf{E_{TP_i}}\cdot \vec{l}$ add the pitch, yaw and squareness errors without and with consideration of the vector of the stylus, respectively.

\subsection{Probing system}
In addition to the deviations resulting from geometric errors the probing process itself is also a source of errors. The VCMM implementation is capable of simulating the systematic and stochastic contributions for contacting probing systems in discrete-point or scanning mode. The simulated probing error $\vec{e}_{P_j}$ for the measurement point $\vec{x}_i$ in the kinematic machine coordinate system is described by the model equation
\begin{eqnarray}
\vec{e}_{P_j} \left(\vec{x}_i\right) = & \left(f_\mathrm{sty_j} e_{\mathrm{Base}_j} + e_\mathrm{{radial}_j} + e_\mathrm{{random}_i} \right)\cdot \vec{n}_i + e_\mathrm{{MP}_j}\cdot \vec{t}_j + \vec{e}_\mathrm{{T}_j} + \mathbf{R}_j\cdot \vec{l}\;,\nonumber
\end{eqnarray}
with $i$ being the index of the measurement point and $j$ indicating the simulation cycle, respectively. The model includes the influences of the stylus $f_\mathrm{sty_j}$
, the base characteristics of the probing system $e_{\mathrm{Base}_j}$, the radius deviations of the probing sphere $e_\mathrm{{radial}_j}$, and a random contribution $e_\mathrm{{random}_i}$ in the direction of the surface normal vector $\vec{n}_i$ of the workpiece. Furthermore, contributions of the multi-probe positioning deviation $e_\mathrm{{MP}_j}$  and the translational and rotational  influences of the exchange of the probe or the probing system $\vec{e}_\mathrm{{T}_j} $ and $\mathbf{R}_j\cdot \vec{l}$ are considered. 

\subsection{Workpiece influence}
The workpiece causes deviations influencing the measurement uncertainty as it expands and shrinks due to the thermal expansion coefficient of the material. Additionally, all real workpieces inherit form deviations, such as surface roughness and waviness. 
\subsubsection{Thermal effects of the workpiece}
Most available CMM are capable of measuring the workpiece's temperature, and thus, use information about the material's coefficient of thermal expansion $\alpha _{\mathrm{W},a}$ to compensate for thermal expansion. However, depending on the quality of the compensation, a residual position deviation $e_\mathrm{P}$ with its Cartesian components
\begin{eqnarray}
e_{\mathrm{P},\{xyz\}} &= p_{\{xyz\}} \left[\alpha _{\mathrm{W},\{xyz\}}\, \Delta T_{\mathrm{W},\{xyz\}} + \Delta \alpha _{\mathrm{W}\{xyz\}}\left(T_{\mathrm{W},\{xyz\}} - T_0\right)\right]\nonumber
\end{eqnarray}
remains uncorrected. On the one hand it depends on the deviation of the workpiece temperature $T_\mathrm{W}$ from the reference temperature of $T_0=20\,^\circ$C and on temperature fluctuations of the temperature measurement system $\Delta T_\mathrm{W}$. On the other hand, additional deviations arise from deviations of the value $\alpha _\mathrm{W}$ from the considered expansion coefficient.

\subsubsection{Form deviations and roughness of the workpiece}
The VCMM considers the contributions of the surface roughness to the measurement uncertainty solely stochastically. The deviation is varied for each individual measurement point independently according to a given probability distribution function. The user is guided by predefined measurement and analysis procedures to help determine the required input parameters. However, to obtain suitable parameters representing the surface properties determination procedures at real surfaces must be performed. Particularly, the effect of filtering depending on the size of the chosen probing sphere plays an important role and has to be considered appropriately. Therefore, resulting uncertainty contributions have to be determined for each combination of probing sphere and surface characteristic individually \citep{keck_et_al2004tm}.
\par
Furthermore, form deviations of the workpiece or of the probed feature contribute to the measurement uncertainty. The strength of their influence depends on the chosen measurement strategy and the geometric specifications of the feature. In general, all features have to be captured entirely in order to obtain a complete picture of the real geometry. However, in practice only a subset of the entire surface can be assessed. Consequently, in this case the measurement uncertainty is underestimated unless the influence of form deviations is considered separately. The VCMM provides the option of simulating the probed features including their oversampling. This is achieved by generating additional measurement points  and profiles to simulate the complete sampling of the feature. In addition, the features are varied by adding synthetic form deviations resulting in different measurement data for the same feature. 
The difference between the results is taken into account for the determined measurement uncertainty.\par
Modeling of realistic form deviations is very costly and only feasible, if the form deviations of the entire workpiece are known. In order to reduce the complexity, the VCMM treats the form deviations in a different manner. In their respective local feature-specific reference coordinate system, the measurement points are distorted using a linear combination of harmonic oscillations, like e.g., 
\begin{eqnarray}
\overline{x}_i = x_i + \sum_{k=1}^{m} A_k\cdot \cos\left(\varphi_k - \frac{k\cdot x_i}{\lambda _k}\right)\,\nonumber
\end{eqnarray}
for the $x$ coordinates. The amplitudes $A_k$ could be extracted from tolerances given by the technical drawings, while for the determination of the wavelengths $\lambda _k$ detailed knowledge about the production process itself should be taken into account. The VCMM varies these parameters and the phase $\varphi_k$ within a wide range of values covering a broad spectrum of possible form deviations. However, it should be noted that the VCMM approach is not meant to estimate the real form deviations, but to provide an additional measurement uncertainty contribution arising from undersampling of surfaces with possible form deviations within a given parameter space. The simulated uncertainty component is decreasing, if the measurement procedure covers the real form deviations more accurately. If the form deviations are completely covered, the analysis using the VCMM simulations does not add any additional uncertainty contribution. In reverse, any undersampling of the feature using only few individual points adds the superposition of the amplitudes to the uncertainty computation.

\subsection{Scanning}
In scanning measurements, the probe is moved along the surface of the workpiece, while constantly touching it, and thus allowing measurement points to be captured continuously. This method enables users to reduce the measurement duration, and is therefore of increasing importance for typical applications in industry. 
However, the procedure of scanning data collection requires consideration of additional, scanning-specific uncertainty contributions.\par
Forces acting during the scanning process may generally alter the measurement points. 
The reasons for the additional contributions are different behavior of the probing system and the dynamic behavior of the machine structure compared to measurements of individual points. There are different ways to implement the additional uncertainty contributions \citep{schwenke_et_al2000DFG_Report}. Extensive studies of state-of-the-art CMM using scanning technology have shown that systematic and random measurement deviations are typically of the same order. Therefore, the VCMM treats uncertainty contributions from tactile scanning as additional random probing uncertainties perpendicular to the workpiece's surface. These stochastic uncertainties can easily be determined empirically for combinations of styli, probing sphere radii, scanning speeds, and probing forces using reference standards. They have to be derived separately 
using both, a calibrated spherical standard and a real surface of the workpiece to be measured.

\subsection{Statistics}
The calculations within in the first version of PTB's VCMM were operated using a fixed number of simulation runs, typically of the order of 200. However, it was observed that the stability of the computed measurement uncertainty was insufficient for some measurement tasks. 
To allow stability testing for all quantities using a single stability threshold value, a normalized stability criterion was introduced into the VCMM:
\begin{eqnarray}
\frac{\Delta s\left(z\right)^2}{s\left(z\right)^2} &< \frac{\delta^\ast}{2} \;, \nonumber\label{eq_stability}
\end{eqnarray}
where $z$ is the mean value of all simulated measurement values, $\Delta s\left(z\right)$ is the variation of the standard deviation of the mean value $ s\left(z\right)$, and $\delta^\ast$ is a dimensionless threshold value chosen by the user.\par 
As soon as the stability criterion (\ref{eq_stability}) is reached, the measurement uncertainty is determined according to the GUM, which also defines how systematic deviations have to be treated and generally assumes that all systematic measurement deviations are corrected. However, in the field of coordinate metrology this requirement cannot be fulfilled at all times due to the economic and/or technical challenges that go along with it. Most often, after applying corrections for probing characteristics and geometry deviations, some residual systematic deviations remain uncorrected, and hence, have to be considered for the combined measurement uncertainty following the applicable guidelines. Therefore, the current VCMM version considers uncorrected systematic uncertainty contributions as well as random ones \citep{haertig_hernla2010VDI}, where the latter are derived from the variance of the simulated measurement values \citep{rost_et_al2016PrecEng}. The combined measurement uncertainty is then calculated as the squared sum of both uncertainty contributions. 

\section{Conclusions and outlook}
The advancement of the existing virtual coordinate measuring machine VCMM enables users to determine the measurement uncertainty for measurements of size, form, and position in discrete-point probing mode and scanning mode, respectively. After successful verification of the simulation technique for scanning measurements used by the VCMM, the software can be used to determine the measurement uncertainty in measurement laboratories and accredited calibration laboratories, respectively. This allows users of industrial measurement facilities to provide measurement uncertainties in conformity assessments and enables accredited calibration service providers to calibrate real workpieces. Furthermore, future extensions of the VCMM will support the use of rotary tables in scanning measurements.\par 



\bibliographystyle{ws-procs9x6} 
\bibliography{LiteraturDH}

\end{document}